\documentclass{knawproc}
\newcommand{\gtorder}{\mathrel{\raise.3ex\hbox{$>$}\mkern-14mu
             \lower0.6ex\hbox{$\sim$}}}
\newcommand{\ltorder}{\mathrel{\raise.3ex\hbox{$<$}\mkern-14mu
             \lower0.6ex\hbox{$\sim$}}}
\newcommand{\proptwid}{\mathrel{\raise.3ex\hbox{$\propto$}\mkern-14mu
             \lower0.6ex\hbox{$\sim$}}}

\begin{document}

\begin{opening}
\title{Young radio galaxies and their environments}
\author{Mitchell C. Begelman$^{1,2}$}
\addresses{%
1. JILA, University of Colorado, Boulder, CO 80309-0440 USA \\
2. Department of Astrophysical and Planetary Sciences, University of
Colorado at Boulder \\
}
\runningtitle{Young radio galaxies}
\runningauthor{Mitchell C. Begelman}
\end{opening}

\begin{abstract}
Many of the powerful radio galaxies observed at high redshift are very
small, presumably because they are very young.  A simple model, which
treats the radio source as a bubble expanding into a radially stratified
medium, is capable of explaining the main features of their size and
luminosity evolution. In particular, the model predicts strong negative
luminosity evolution with increasing size, thus accounting for the large
number of
short-lived small sources in flux-limited samples.

A variety of observational characteristics can be understood within the
framework of the bubble model.  I address three of these: 1) the number
versus size statistics, which may provide a clue that jets in radio
galaxies are intermittent, with ``on--off" cycles lasting only
$10^4-10^5$ yr; 2) the rarity of spectral steepening due to synchrotron
cooling at frequencies of a few GHz, and its implications for source
geometry and assumptions about equipartition; and 3) the prevalence of
gigahertz-peaked spectra among small sources.

I argue that all of these phenomena are primarily the consequence of the
youth and compactness of the radio source, and depend only weakly on
environment.  There is no need to postulate an ISM with properties any
different from that of an ordinary nearby galaxy.

\end{abstract}

\section{Introduction}

Powerful double radio sources are observed to range in projected linear
size ($LS$) from less than 100 pc to more than a megaparsec. They have
been divided arbitrarily into a sequence of size categories, ranging
from compact symmetric objects (CSOs, $LS < 500$ pc: Wilkinson et al.
1994), to medium symmetric objects (MSOs, 500 pc $< LS < 15$ kpc: Fanti
et al. 1995), to
full-size FR II radio galaxies ($LS > 15$ kpc).  The CSOs contain two
important subcategories  based on spectral classification, compact steep
spectrum (CSS) sources (Fanti et al. 1990) and gigahertz-peaked spectrum
(GPS) sources (O'Dea, Baum, \& Stanghellini 1991).  There is still some
debate about whether the small sources are young versions of the large
sources (see, e.g., O'Dea \& Baum 1997).  Alternative views propose that
the small sources are short-lived or evolve into some other type of
source (e.g., FR I radio galaxies) as they grow.  In this paper I will
adopt the view that all of these categories fit into a single
evolutionary sequence, which I believe is the easiest hypothesis to
reconcile with both observations and theory.

According to the evolutionary scenario, the small sources are plentiful
not because their growth is stunted by interaction with an exceptionally
dense interstellar medium (ISM), or because the average source is
short-lived.  Rather, the brevity of the compact phase is partly
compensated by strong negative luminosity evolution (Readhead et al.
1996), which allows the smaller sources to be included in flux-limited
surveys out to much larger distances. This luminosity evolution is
expected on the basis of simple physical models (Begelman 1996).

To capture the main features of radio source evolution, it suffices to
treat the cocoon as an adiabatic spherical bubble, filled with
relativistic fluid.  Real cocoons are elongated along the jet axis, but
the ``aspect ratio" (length/width) is seldom much larger than 3,
presumably because the ``dentist's drill" effect (Scheuer 1982) spreads
out the ram pressure of the jets over a large solid angle. The cocoon
and shell of shocked ISM expand supersonically into the ambient medium
at a speed $v \sim [p_c (t)/ \rho(r)]^{1/2}$, where $p_c (t)$ is the
cocoon pressure at time $t$ and $\rho (r)$ is the ambient density at the
radius of the outer shock.
The synchrotron emissivity comes mainly from within the cocoon, and is
probably dominated by ``lobe" regions near the ends of the jets, where
the pressure is a few times the mean cocoon pressure (see section 3).
Significant synchrotron emission may also come from the hotspots,
instantaneous impact points of the jets, where the pressure can be much
larger still.

A useful family of models can be derived by assuming a power-law
distribution of ambient density, $\rho(r) \propto r^{-\delta}$.  The
internal structure of the cocoon is assumed to evolve self-similarly
(Falle 1991; Begelman 1996; Kaiser \& Alexander 1997), with a
relativistic electron energy distribution of fixed slope and a fixed
ratio of magnetic to relativistic particle energy density,
$\varepsilon_{\rm mag} / \varepsilon_{\rm rel} \equiv \zeta \leq 1$.
The latter parameterization serves to generalize the usual assumption of
equipartition ($\zeta = 1$). We consider $\zeta < 1$ since synchrotron
cooling is less efficient in this regime, a desirable feature for
compact sources as we shall see in section 3.

Using this simple model, one can explain the observed number versus size
relation $N \proptwid (LS)^{0.4}$ (Fanti et al. 1995; Readhead et al.
1996), which seems to apply over limited ranges of $LS$, if the density
index $\delta$ lies in the range 1.5--2 (Begelman 1996). The velocity of
expansion is nearly constant and the radio power (at a fixed frequency)
declines roughly as $(LS)^{-1/2}$. For ISM number density $n(r) = n_0
R_{\rm kpc}^{-2}$ cm$^{-3}$ (where $R_{\rm kpc} = r / 1$ kpc) and jet
power $L_j = 10^{45} L_{45}$ erg s$^{-1}$, we obtain the following
estimates of the cocoon pressure, expansion speed, and elapsed lifetime
for a source that has expanded to radius $r$:
\begin{eqnarray}
p_c &=& 6 \times 10^{-8}\  n_0^{1/3} L_{45}^{2/3} R_{\rm kpc}^{-2} \;
{\rm dyn \ cm}^{-3} \\
v &=& 2700\ n_0^{-1/3} L_{45}^{1/3} \;   {\rm km \ s}^{-1} \\
t &=& 0.4\ n_0^{1/3} L_{45}^{-1/3} R_{\rm kpc} \;   {\rm Myr}.
\end{eqnarray}

The ``bubble model" seems to provide a robust framework for
understanding both the luminosity and size evolution of young radio
galaxies.  In this paper, I will show that it also readily accommodates
the next level of observational detail.  In section 2, I show that the
``plateau" in the number--size relation, discovered by O'Dea \& Baum
(1997), can be interpreted as the consequence of intermittency, with the
jets switching on and off on timescales of $10^4-10^5$ yr.   In section
3 I analyze the synchrotron emissivities predicted by the bubble model.
Spectral steepening at a few GHz, attributable to synchrotron cooling,
is seldom observed.  Given the high synchrotron efficiencies predicted
by the bubble model for small sources, the scarcity of evidence for
cooling suggests that most of the emission comes from lobe regions only
briefly traversed by the radiating electrons.  It may also indicate the
presence of
sub-equipartition magnetic fields. Finally, I examine the possible
causes of gigahertz-peaked spectra in section 4.  As an alternative to
extant models invoking synchrotron self-absorption or free-free
absorption by a screen, I propose a model in which the free-free
absorption is produced by interstellar clouds that have been engulfed by
the bubble, and whose surface layers are photoionized by UV radiation
internal to the bubble.  This model naturally predicts the inverse
correlation between turnover frequency and source size, as well as the
typical change in spectral index at the turnover, without any special
assumptions about the ISM density.

Except where otherwise noted,  all succeeding calculations are based on
the $\rho \propto R^{-2}$ model described in equations (1)--(3).  The
reader should keep in mind that the results presented may depend
quantitatively on $\delta$, although the qualitative dependence (for
$1.5 \leq \delta \leq 2$) will be weak.

\section{Do source counts imply intermittency?}

A power-law relationship between number counts and size does not seem to
continue uniformly across the entire range of source sizes. When O'Dea
\& Baum (1997) combined data on the two principal classes of CSOs---the
gigahertz-peaked spectrum (GPS) and compact steep spectrum (CSS)
sources---with data on 3CR classical doubles, they found the
$(LS)^{0.4}$ behavior above 10 kpc, and perhaps some hint of a similar
behavior at $LS \ltorder 0.3$ kpc.  But between 0.3 and 10 kpc, $N$ is
nearly independent of $LS$. They suggested that GPS and CSS sources
might be overabundant, relative to the extrapolation from larger
sources, because they are
short-lived and never evolve into large sources, because their growth is
``frustrated" by an exceptionally dense ISM, or because they decline in
luminosity more rapidly than predicted by the bubble model.  Since an
explanation in terms of environmental differences is not attractive due
to the insensitivity of bubble evolution to ambient density, their
interpretation would suggest that the GPS--CSS sources form a distinct
class of objects.

However, Reynolds \& Begelman (1997) showed that a ``plateau" in the
$N-LS$ distribution could simply indicate that radio galaxies are
intermittent.  According to the bubble model, the cocoons of small
sources are highly overpressured with respect to the ambient medium.
They would therefore keep expanding supersonically long after the
central source turned off. The  radio emission, dominated by the regions
near the hotspots, would fade quickly. Once the nucleus turned on again,
the cocoon would partially repressurize, the radio flux would rise, but
the emissivity would be lower than before due to the increased volume
and lower pressure.  If a large source of instantaneous power $L_j$ had
experienced many on--off cycles, its evolutionary state would be
indistinguishable from that of a source that had been on all the time,
but had a {\it mean} power of $(t_{\rm on} / t_{\rm off} + t_{\rm on})
L_j $. Reynolds \& Begelman (1997) found that the factor $\sim 10$
difference between the actual
GPS--CSS number counts and the extrapolation of 3CR number counts could
imply that radio galaxies are ``on" only about 30\% of the time.  (These
calculations assumed $\delta = 1.8$.) Ages corresponding to the source
sizes bounding the plateau measure the typical durations of the ``on"
and ``off" cycles.  These are surprisingly short, of order 30,000 and
70,000 yr, respectively.  The timescales are intriguingly close to the
timescales of viscous accretion disk instabilities studied by
Siemiginowska \& Elvis (1997).

A prediction of the intermittency hypothesis is that there should be
$\sim 3$ times as many radio galaxies in the ``off" state as in the
``on" state.  These might be detectable in low-frequency, low-surface
brightness radio surveys, since their spectra would tend to be steep and
their emission diffuse. One might also search for X-ray emission from
the shell of shocked ISM that surrounds the cocoon (Heinz, Reynolds, \&
Begelman 1997).  Whereas the radio emissivity probes the instantaneous
jet power, the X-ray emission is more indicative of the mean jet power
averaged over the lifetime of the source. By comparing the radio and
X-ray properties of Perseus A (NGC 1275), Heinz et al. (1997) deduced
that Per A might currently be in an ``off" state.

\section{Synchrotron emissivity and radiative efficiency}

Young, compact, and therefore high-pressured sources should radiate
synchrotron emission much more efficiently than their larger
descendants.  Where the synchrotron cooling timescale of radio-emitting
electrons is shorter than the expansion timescale (assumed to be of the
same order as the timescale for injecting freshly accelerated
electrons), the spectral index should steepen by $\Delta\alpha \approx
0.5$ over the value determined by the particle acceleration process.
While compact steep spectrum (CSS) sources all have spectra steeper than
0.5, they are not often observed with $\alpha > 1$ at GHz frequencies,
as would be expected if cooling were important.

The lack of spectral evidence for cooling places a significant
constraint on the region that dominates the emission.  If the observed
synchrotron radiation came from the bulk of the cocoon, the bubble model
would predict steepening above frequencies
\begin{equation}
\nu_{\rm cool} \sim 3.5\times 10^{-3} \ \zeta^{-3/2} n_0^{-7/6}
L_{45}^{-1/3} R_{\rm kpc} \; {\rm GHz}.
\end{equation}
The monochromatic radio power at 5 GHz is given by
\begin{equation}
P_5 \sim 3\times 10^{28} \ \zeta^{3/4} n_0^{7/12} L_{45}^{7/6} R_{\rm
kpc}^{-1/2} \; {\rm W \ Hz}^{-1}.
\end{equation}
Thus, for observed sources with a typical 5 GHz power $10^{27} P_{27}$ W
Hz$^{-1}$, we predict that the spectrum should steepen above
\begin{equation}
\nu_{\rm cool} \sim 9 \ P_{27}^{-2/7} \zeta^{-9/7} n_0^{-1} R_{\rm
kpc}^{6/7} \; {\rm MHz}.
\end{equation}
In other words, if the mean emissivity of the cocoons dominated the 5
GHz emission then all CSOs should have $\alpha \gtorder 1$ unless $n_0
\zeta^{9/7} < 10^{-3}$. The latter exception involves such extreme
conditions (low ambient densities and/or sub-equipartition magnetic
fields) that it is unlikely to be satisfied.

This situation is ameliorated somewhat if the emission comes mainly from
the overpressured lobes within the cocoon. The important effect is not
the higher emissivity in the lobes (the high emissivity in the cocoon is
the cause of the rapid cooling problem), but rather the fact that the
electrons spend a short time in the high-pressure region, compared to
the age of the source.  Suppose that the lobes occupy a fraction $x$ of
the cocoon's volume and that the lobe pressure is $y > 1$ times the
cocoon pressure. In order for the lobes to dominate the synchrotron flux
(if $\zeta$ has the same value throughout) we must have $x y^{7/4} > 1$.
If the relativistic electrons traverse the lobes at a speed $v_l \sim
0.3 c$ then the steepening occurs at
\begin{equation}
\nu_{\rm cool} \sim 70 \ x^{4/21} P_{27}^{-6/7} (v_l/ 0.3 \ c)^2
\zeta^{-6/7} R_{\rm kpc}^{4/7} \; {\rm GHz}.
\end{equation}
Note that equation (7) depends weakly on $x$ and not at all on $y$ or
$n_0$.  Although this result seems more consistent with observations
than equation (6), it is sensitive to the highly uncertain value of
$v_l$, which may well be smaller than our optimistic fiducial value. If
this analysis is correct, we might expect the spectra of most CSOs to
steepen at frequencies $\ltorder 100$ GHz, and to find an inverse
correlation between 5 GHz power and steepening frequency. If the spectra
of the most powerful CSOs do not break at high frequency, it could
indicate that the magnetic fields are below equipartition.

\section{Absorption in gigahertz-peaked sources}

It is not yet certain whether the spectral turnovers in GPS sources are
caused by synchrotron self-absorption (O'Dea \& Baum 1997), free-free
absorption (Bicknell, Dopita, \& O'Dea 1997), or some other mechanism
(G. Bicknell 1997, private communication).  Two features which should be
explained be any successful model are the typical change in spectral
index across the turnover, $\Delta \alpha \sim 2 \pm 0.5$, and the
relationship between turnover frequency and source size, $\nu_t \sim
R_{\rm kpc}^{-0.7} $ GHz (O'Dea \& Baum 1997).

If the bulk of the cocoon dominated the emission, synchrotron
self-absorption could be ruled out immediately as the main absorption
mechanism.  The brightness temperature at 5 GHz would then be  $T_{b, 5}
\sim 5\times 10^6 \ P_{27} R_{\rm kpc}^{-2}$ K, orders of magnitude
below the energies of synchrotron emitting electrons.  However, as we
saw above, the emission is likely to be dominated by high-surface
brightness compact regions within the cocoon. If we restrict the
emission to the overpressured lobes, parameterized as above, the
brightness temperature increases by a factor $x^{-2/3}$. To reach the
brightness temperature required for
self-absorption at 5 GHz, $T_{b,5}\sim 6 \times 10^{12} \ x^{1/7}
P_{27}^{-1/7} \zeta^{-1/7} R_{\rm kpc}^{3/7}$ K, the lobes would have to
subtend an angle on the sky less than one percent that of the entire
source, $x^{1/3} \ltorder 0.003 \ P_{27}^{-8/17} \zeta^{3/17} R_{\rm
kpc}^{-1}$.  Note that the required size of the emitting region is
completely independent of the pressure enhancement in the lobes (the
factor $y$). Such a compact emitting region seems implausible, given
what we know about the morphologies of large FR II radio galaxies; in
any case, the prediction could be tested through VLBI observations.

Free-free absorption seems a more promising mechanism for shaping the
spectral peaks of GPS sources.  In modeling the absorption there is a
choice of geometries, ranging from an external screen to a spray of
absorbing clouds mingled with the synchrotron-emitting gas.  Bicknell et
al. (1997) have adopted the former approach, attributing the absorption
to the shell of shocked ISM.  In order to obtain the observed change of
spectral slope across the GHz-peak, rather than an exponential cutoff,
they argue for a particular power-law distribution of column densities
covering the source.  In order to obtain enough emission measure in the
absorbing layer, the shock has to be able to cool in less than the
expansion timescale of the cocoon.  This, plus the requirement that the
shock be autoionizing (i.e., that the cooling layer emit enough UV
radiation to ionize the surrounding gas), forces one to demand a rather
low shock speed ($\ltorder 1000$ km s$^{-1}$) and high ambient density
(10--100 cm$^{-3}$).

\subsection{The ``engulfed cloud" model}

Constraints on the ambient conditions are relaxed considerably if the
absorption is supplied by interstellar clouds engulfed by the cocoon.
The ``engulfed cloud"  model has the advantage that it automatically
yields $\Delta \alpha \sim 2$ across the spectral peak, due to the fact
that the free-free opacity varies with frequency $\proptwid \nu^{-2}$.
An important difference between the Bicknell et al. model and the
engulfed cloud model is the source of the ionizing radiation.  Bicknell
et al. rely on radiation from the narrow layer of cooling shocked gas to
ionize the (much thicker) neighboring layers of gas that have already
cooled.  In the engulfed cloud model, a ``skin" on the surface of each
cloud is photoionized by a combination of synchrotron UV emission from
the lobes (if relativistic electrons are accelerated to high enough
energies in the hotspots) and radiation from the active nucleus.  If
clouds at $10^4 T_4$ K are in pressure equilibrium with the interior of
the cocoon and are irradiated by $10^{45} L_{UV, 45}$ erg s$^{-1}$ in
diffuse UV luminosity, the neutral fraction in the ionized layer is
given by $n_H / n_e \sim 0.1 \ T_4^{-3/4} n_0^{1/6} L_{UV, 45}^{-1/2}
L_{45}^{1/3}$ and is independent of $R$.  In the overpressured lobes,
this ratio is increased by a factor of $y^{1/2}$.  Thus, for a wide
range of reasonable parameters we can consider the layer to be fully
ionized. If the column density is sufficiently large that the clouds are
ionization bounded, the emission measure of the photoionized gas
saturates at
\begin{equation}
EM_{\rm max}  = \langle n_e^2 R \rangle_{\rm max} \sim 3 \times 10^{24}
\ T_4^{1/2} L_{UV, 45} R_{\rm kpc}^{-2} \; {\rm cm}^{-5}
\end{equation}
and the turnover frequency, which scales as $(EM)^{1/2}$, is given by
\begin{equation}
\nu_t \sim 0.6 \ T_4^{-0.4} L_{UV, 45}^{1/2} R_{\rm kpc}^{-1} \; {\rm
GHz}.
\end{equation}
Given the crudeness of the assumptions that went into its derivation,
equation (9) agrees remarkably well with the observational result $\nu_t
\sim R_{\rm kpc}^{-0.7} $ GHz obtained by O'Dea \& Baum (1997).

It does not take much entrained matter to produce the observed
absorption in the engulfed cloud model.  The maximum filling factor of
ionized gas is
\begin{equation}
f = { EM_{\rm max} \over n_e^2 R } \sim 5 \times 10^{-7} \ T_4^{5/2}
n_0^{-2/3} L_{UV, 45} L_{45}^{-4/3}  R_{\rm kpc}^2,
\end{equation}
corresponding to a {\it mean} density
\begin{equation}
\langle n_e \rangle = f n_e \sim  0.02 \ T_4^{3/2} n_0^{-1/3} L_{UV, 45}
L_{45}^{-2/3}  \; {\rm cm}^{-3}.
\end{equation}
This is smaller than the amount of cold matter likely to be present in
the ambient ISM, and validates the assumption that the clouds are
ionization bounded and that the emission measure is close to the maximum
possible value.  Indeed, the amount of ionized gas is far smaller than
the mean density needed to confine the bubble as it passes through its
GPS phase.  This implies either that most of the matter within and
around the bubble is neutral or that the ISM mass is dominated by the
diffuse intercloud medium.

The small amount of matter required for GPS absorption is also
reassuring from a hydrodynamical perspective.  Simulations by Klein,
McKee, \& Colella (1994) suggest that cold clouds engulfed by a blast
wave can be shredded in several times the ``cloud-crushing" timescale.
The latter is given by $\chi^{1/2} \equiv (\rho_{\rm cloud} / \rho_{\rm
ambient})^{1/2}$ times the
sound-crossing time across the cloud.  If the shocked ISM shell has a
thickness about 10\% the radius of the bubble, and the cloud density in
the ISM is $\chi_0 \sim 100$ times that of the intercloud medium, then
clouds should be able to survive passage through the shocked shell if
the cloud size satisfies $r_{\rm cl} \gtorder 0.01 \ \chi_{100}^{-1/2}
T_4^{1/2} L_{45}^{-1/3} n_0^{1/3} R_{\rm kpc}$ pc, where we have taken
the cloud destruction time to be 4 times the crushing time.  Since the
mean density of the hot gas inside the bubble is likely to be more than
100 times lower than that of the shocked shell (Bicknell \& Begelman
1996), clouds which survive the shell-crossing are likely to survive
once inside the bubble.  But it is not certain whether such large clouds
will have enough covering factor to blanket the entire synchrotron
source.  This is where the modest mass requirements of the engulfed
cloud model can work to our advantage, since there may be adequate
opacity (with high covering factor) in the residual shredded clouds,
even if the bulk of the swept-up cloud mass has been mixed into the
low-density hot phase.  We also note that the smaller swept-up clouds
might never penetrate  the cocoon, but rather join the outward motion of
the ISM shell.  The morphology of the
line-emitting gas surrounding the M87 bubble suggests that much of this
material has been excluded from the cocoon (Bicknell \& Begelman 1996).
Even if the clouds pile up on the outer surface of the cocoon, however,
ionization by UV from inside the cocoon could still produce an absorbing
layer with similar properties to those predicted by the engulfed cloud
model, at the  cost of sacrificing a natural explanation for
$\Delta\alpha \sim 2$.

We conclude by noting that the engulfed cloud model, as well as the
Bicknell et al. (1997) screen model, can easily account for the observed
Faraday rotation measures in the range $10^3-10^4$ rad m$^{-2}$.  Since
the depth to which one can ``see" in the cloud model scales as $\nu^2$,
the effective rotation measure scales similarly at frequencies below the
turnover.  This implies that the Faraday rotation angle (or dispersion)
should be independent of frequency below the turnover, while displaying
the normal $\nu^{-2}$ behavior at higher frequencies.  If this effect
could be detected observationally, it could be used to distinguish
between the engulfed cloud model and screen models, since in the latter
the Faraday rotation scales as $\nu^2$ on both sides of the turnover.

\section{Conclusions}

I have tried to show how readily the simple bubble model of young radio
galaxies can accommodate a variety of observational details, with few
additional assumptions.  The existence of a ``plateau" in the number
versus size counts plotted by O'Dea \& Baum (1997) need not imply
physically distinct classes of source.  Instead, it could constitute
evidence for intermittency in the central engines of powerful radio
galaxies (Reynolds \& Begelman 1997). The lack of steepening
attributable to synchtron cooling at a few GHz is harder to explain, as
small radio sources should be extremely efficient synchrotron emitters.
An observable cooling signature can be avoided, at frequencies $\ltorder
100$ GHz, if the emission is dominated by electrons traversing the
overpressured lobe regions of the cocoon, as indeed is the case in large
FR II sources.  Magnetic fields of sub-equipartition strength can also
help.

There are several possible explanations for the spectral turnovers in
GPS sources.  Synchrotron self-absorption is the least viable, since it
would require most of the emission to come from extremely tiny spots
within the cocoon.  Free-free absorption, on the other hand, is
extremely promising, and may arise in a foreground screen and/or
embedded clouds within the cocoon.  To provide an alternative to the
Bicknell et al. (1997) screen model, I proposed a simple model in which
the absorption arises in interstellar clouds that have been engulfed by
the cocoon. The compressed surface layers are kept photoionized by UV
from the central engine and the relativistic cocoon plasma itself.  The
emission measure in the ionization bounded clouds saturates at a value
that approximates the observed relationship between turnover frequency
and source size, and the model naturally reproduces the typical change
in spectral index across the turnover, $\Delta\alpha \sim 2$.  I
suggested a diagnostic, involving the Faraday rotation as a function of
frequency near the turnover, that might help to distinguish between
screen and embedded cloud models.

With the exception of the high densities required by the Bicknell et al.
(1997) GPS model, none of the distinctive properties of young, compact
radio galaxies discussed above require ambient conditions that are any
different from those found in ordinary, nearby galaxies. Most of the
relations I have derived depend weakly on the ambient ISM density, so we
cannot use the properties of CSOs to argue that galaxies at high-$z$ had
ISMs similar to those we find today.  What we can conclude is that the
distinctive properties of young radio galaxies reflect their relative
youth much more sensitively than they reflect their environments.

\begin{acknow}
This work has been supported by the National Science Foundation under
grant AST-9529175.  I am grateful to Chris Reynolds and Sebastian Heinz
for their valuable insights.
\end{acknow}

\end{document}